\documentclass[english,aps,floats,twocolumn,showpacs,nofootinbib]{revtex4}
\usepackage{pslatex}
\usepackage[T1]{fontenc}
\usepackage[latin1]{inputenc}
\usepackage{graphicx}
\usepackage{epsfig}

\usepackage{calc}
\usepackage{ifthen}

{\newcommand{\lsim}{\mbox{\raisebox{-.6ex}{~$\stackrel{<}{\sim}$~}}}
{\newcommand{\gsim}{\mbox{\raisebox{-.6ex}{~$\stackrel{>}{\sim}$~}}} 
{\newcommand{\Rb}{R\hskip-1.5mm\raisebox{-1.00mm}/_{p}}
\newcommand{\bea}{\begin{eqnarray}}
\newcommand{\eea}{\end{eqnarray}}
\newcommand{\Mchitilde}{M_{\overline{\chi}}}

\newcommand{\MLtilde}{M_{\tilde{L}_i}}

\newcommand{\nchitilde}{n_{\overline{\chi}}}
\newcommand{\omegachitilde}{\Omega_{\overline{\chi}}}

\newcommand{\nc}{\newcommand}
\nc{\renc}{\renewcommand}
\nc{\eqs}[2]{\mbox{Eqs.~(\ref{#1},\,\ref{#2})}}
\nc{\eq}[1]{\mbox{Eq.~(\ref{#1})}}
\nc{\figs}[2]{\mbox{Figs.~(\ref{#1},\,\ref{#2})}}
\nc{\fig}[1]{\mbox{Fig~.(\ref{#1})}}
\nc{\be}[1]{\begin{equation} \mbox{$\label{#1}$}}
\nc{\ee}{\vspace{0.1cm}\end{equation}}

\newcommand{\bean}{\begin{eqnarray*}}
\newcommand{\eean}{\end{eqnarray*}}

\def\GeV{{\rm \ GeV}}

\def\TeV{{\rm \ TeV}}

\begin{document}
\title{$Z_2$-Singlino Dark Matter in a Portal-Like Extension of the Minimal Supersymmetric Standard Model}
\author{John McDonald}
\email{j.mcdonald@lancaster.ac.uk}
\author{Narendra Sahu}
\email{n.sahu@lancaster.ac.uk}
\affiliation{Cosmology and Astroparticle Physics Group, University of 
Lancaster, Lancaster LA1 4YB, UK}
\begin{abstract}

    We propose a $Z_2$-stabilised {\it singlino} ($\overline{\chi}$) as a dark 
matter candidate in extended and $R$-parity violating versions of the supersymmetric 
standard model. $\overline{\chi}$ interacts with visible matter via a heavy
messenger field $S$, which 
results in a supersymmetric version of the Higgs portal interaction. The relic abundance of $\overline{\chi}$
can account for cold dark matter if the messenger mass satisfies $M_S \lsim 10^{4}$ GeV. Our 
model can be implemented in many realistic supersymmetric models such as the NMSSM and nMSSM.

\end{abstract}
\pacs{12.60.Jv, 98.80.Cq, 95.35.+d}
\maketitle
\section{Introduction} 
It is well established that visible matter is not sufficient to account for the observed 
structure of the Universe. This implies the existence of non-baryonic dark matter (DM).
Global fits of cosmological parameters can accurately determine the density of DM, 
albeit indirectly. Measurements of the anisotropy of the cosmic microwave background radiation (CMBR) and 
of the spatial distribution of galaxies give for the density of DM~\cite{pdg}
\begin{equation}
\Omega_{\rm DM}h^2=0.106\pm 0.008\,.
\end{equation}  
Identifying the nature of dark matter is a major goal in 
astroparticle physics. Many particle physics candidates have been proposed
in both supersymmetric (SUSY) and non-supersymmetric extensions of the standard 
model (SM). In either case the stability of DM is ensured by imposing a global 
symmetry. The simplest global symmetries considered are $Z_2$ and $U(1)$; see for 
example \cite{u1group,z2group,singlets,singlet0,singlet2,inert}.  

In low energy effective SUSY theories the symmetry is usually $R$-parity,
$(-1)^{(3B+L+2S)}$, which is imposed to  conserve baryon (B) and lepton (L) numbers. 
As a result the stability of proton is ensured. It turns out that $R$ is +1 for 
all SM fields and -1 for their superpartners. Thus $R$-parity, which is a $Z_2$ 
symmetry, protects the decay of lightest SUSY particle (LSP) to SM particles. As a 
result the LSP is a good candidate for DM within minimal SUSY standard model (MSSM) 
and its extensions as long as the conservation of $R$-parity is ensured.   

  However, B and L are accidental global symmetries of SM. Thus it is not clear {\it a priori} 
that B and L are conserved within the MSSM. If B and L are violated then $R$-parity is 
not conserved. Non-conservation of $R$-parity is one way to generate small neutrino 
masses~\cite{numass_group}, which provide solid evidence for phyiscs beyond the SM. Moreover, if 
$R$-parity is violated then leptogenesis is possible~\cite{rparitylep_group}, which 
explains the small matter anti-matter asymmetry (${\cal O}(10^{-10})$) required for 
successful Big-Bang nucleosynthesis. However, within the MSSM and its extensions there 
is no well-motivated particle physics candidate for DM in the presence of $R$-parity 
violation\footnote{In supergravity (SUGRA) theories, the gravitino can account for dark 
matter in certain regions of parameter space since its coupling with matter fields is 
suppressed by the Planck scale~\cite{gravitinodarkmatter}.}.

               In the following we will explore an alternative possibility for the DM 
candidate in SUSY models, irrespective of whether $R$-parity is violated or conserved, by 
introducing a new $Z_{2}$ symmetry and additional singlet fields. Singlet extensions of the MSSM 
are often considered to ensure that the $\mu$ parameter is at the electroweak scale~\cite{kim&nilles}. The prime among them are the NMSSM (the Next-to-Minimal SUSY 
Standard Model) and the nMSSM (the nearly-Minimal SUSY Standard Model). In such models, if $R$-parity 
is conserved then the DM candidate can be an $R$-parity odd singlino ~\cite{singlino_group}. 
Here we propose an alternative SUSY DM candidate: a $Z_2$-odd singlino ($\overline{\chi}$) which 
is stable without requiring $R$-parity\footnote{A different $Z_{2}$-singlino dark matter model, which is based on a broken $U(1)$ gauge group, was presented in \cite{lee}.}.

   Beyond considerations of $R$-parity, $Z_2$-singlino dark matter is interesting as a SUSY 
implementation of gauge singlet dark matter. Gauge singlet scalar dark matter interacting via 
the Higgs portal \cite{portal} was first discussed in detail in \cite{singlets}, with a further 
study presented in \cite{singlet2} and an earlier analysis given in \cite{singlet0}. With the 
advent of the LHC, Higgs portal couplings to hidden sector particles are of considerable 
topical interest. The superpotential coupling we will consider here is the natural extension 
to SUSY of the Higgs portal concept. However, it is necessarily non-renormalisable are a 
consequence of SUSY, pointing to the existence of further new particles at the TeV scale.

\section{Model for $Z_{2}$-Singlino Dark Matter}

\subsection{R-parity conserving SUSY}

We extend the MSSM by adding a chiral superfield $\chi$ and a messenger 
field $S$. We also impose an additional $Z_2$ symmetry under 
which $\chi$ is odd, while all other fields are even. The full superpotential is 
\be{j1} W = W_{MSSM} + \lambda_{1} S \chi \chi + \lambda_{2} S H_{u} H_{d} + \frac{M_{S}}{2} S^2 
+ \frac{M_{\chi}}{2} \chi^2     ~,\ee
where
\begin{equation}
W_{\rm MSSM}=h^e_{ij} L_i \ell^c_j H_d+h^u_{ij} Q_i u^c H_u +
h^d_{ij} Q_i d^c H_d +\mu H_u H_d \,.
\label{mssm-sup}
\end{equation}
In this case 
the effective superpotential after integrating out $S$ becomes 
\begin{equation}
W =W_{\rm MSSM} + \frac{M_\chi}{2} \chi^2 + \frac{f \;\chi^2 H_u
H_d}{M_S} 
~,\end{equation}
where $f = \lambda_{1} \lambda_{2}$.  
The term with coupling $f$ is the natural generalisation to SUSY of the Higgs portal-type 
coupling to $\chi$ scalars of the form $\chi^{\dagger} \chi H^{\dagger} H$ \cite{portal}. 
However, SUSY implies that the Higgs portal interaction is now non-renormalisable.
The Lagrangian terms involving the interaction of $\chi$ scalars and fermions, 
to order $1/M_S$, are then
\begin{eqnarray}
-\mathcal{L_\chi} &\supset&  |M_\chi|^2 \chi^\dagger \chi + M_\chi \overline{\chi}.
\overline{\chi}+   \left[      \frac{f \; M_{\chi}}{M_S} \chi \chi^\dagger H_u H_d \nonumber \right.\\ 
&& \left. + \frac{f}{M_S} \chi^2 \overline{H}_u.\overline{H}_d + \frac{f}{M_S}\chi H_d \overline{\chi}. 
\overline{H}_u + \frac{f}{M_S}\chi H_u \overline{\chi}. \overline{H}_d \nonumber \right.\\ 
&& \left. + \frac{f}{M_S} H_u H_d \overline{\chi}.\overline{\chi} + {\rm h.c.}  \right] + O(1/M_S^2)\,,
\end{eqnarray}
where $\chi$ denotes the scalar and $\overline{\chi}$ the two-component fermion.

\subsection{$R$-parity violating SUSY}
The superpotential involving $R$-parity non-conserving interactions is:
\begin{equation}
W \supset W_{\Rb} + \frac{M_\chi}{2} \chi^2 + h_i \frac{\chi^2 
L_i H_u}{M_S}\,,
\label{R-violating}
\end{equation}
where
\begin{equation}
W_{\Rb}=\lambda_{ijk}L_i L_j \ell^c_k + \lambda^{'}_{ijk}L_i 
Q_j d^c_k +\lambda^{''}_{ijk}u^c_i d^c_j d^c_k
+\mu_i^{'} L_i H_u\,
\label{rp-sup}
\end{equation}
is the $R$-parity non-conserving superpotential in MSSM. The $R$-parity violating 
terms in the Lagrangian involving the interaction of $\chi$,  to order $1/M_S$, 
are then given by
\begin{eqnarray}
-\mathcal{L_\chi} &\supset&  |M_\chi|^2 \chi^\dagger \chi + M_\chi \overline{\chi}.
\overline{\chi} +  \left[ \frac{h_i \; M_\chi}{M_S} \chi \chi^\dagger \tilde{L}_i H_u \right. \nonumber\\
&& \left. + \frac{h_i }{M_S}\chi^2 \overline{L}_i .\overline{H}_u +  \frac{h_i}{M_S}\chi H_u \overline{\chi}.\overline{L}_i 
+\frac{h_i }{M_S}\chi \tilde{L}_i \overline{\chi}.\overline{H}_u \right. \nonumber\\
&& \left. +  \frac{h_i}{M_S} \tilde{L}_i H_u \overline{\chi}.\overline{\chi} + {\rm h.c.} \right] 
+ O(1/M_S^2)   ~,
\end{eqnarray}
where $\tilde{L}_{i}$ is the slepton doublet. 

\subsection{Gauge singlet dark matter} 
Both the scalar and fermion components of the $\chi$ superfield are stable due to the 
$Z_{2}$ symmetry and therefore the lightest of these will be a potential DM candidate. In most cases 
the lightest component will be the fermion, the $Z_{2}$-singlino $\overline{\chi}$, since the scalar 
component will gain additional mass from SUSY breaking. We will therefore focus on the 
$Z_{2}$-singlino as the DM candidate \footnote{There may be regions of parameter space where 
the SUSY mass $M_{\chi}$ is close to the SUSY breaking mass terms, in which case the scalar 
$\chi$ could be the lightest component. We will return to this case in future work.}.    
Its relic abundance will then be determined by the following scattering processes:
\begin{eqnarray}
\overline{\chi} \overline{\chi} & \rightarrow & {\rm MSSM \; fields}\nonumber\\
\overline{\chi} \overline{\chi} &\rightarrow & \chi^\dagger \chi\nonumber\\
\overline{\chi} \chi  &\rightarrow & {\rm MSSM \;  fields} \,.
\end{eqnarray}
The latter two processes will be negligible due to Boltzmann suppression if the $\chi$ mass 
is large compared with the $\overline{\chi}$ mass. We will assume this to be the case in the 
following. Therefore we will only consider the first class of processes when calculating the 
relic abundance of $\overline{\chi}$.

\section{Relic Abundance of $Z_{2}$-Singlinos}
In this section we calculate the relic abundance of $\overline{\chi}$. 
We first calculate the scattering cross-section times relative velocity for annihilation 
processes to MSSM final states. 

     After electroweak symmetry breaking there are five physical Higgs scalar degrees of freedom. In this letter we will consider the physical Higgs scalars to correspond to gauge eigenstates when calculating the cross-sections, with all Goldstone bosons coming from $H_{d}$. The physical Higgs scalars are assumed to have a common mass $M_{H}$. In addition, we will consider the gaugino and Higgsino gauge eigenstates to correspond to mass eigenstates with a common neutralino mass. A more general analysis will be presented in future work. 

\begin{figure}[htbp]
\begin{center}
\epsfig{file=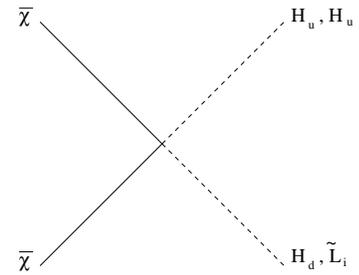, width=0.25\textwidth}
\caption{The four point annihilation of $\overline{\chi}\overline{\chi}$ to Higgs and 
sleptons in MSSM}
\label{fig1}
\end{center}
\end{figure}
In the non-relativistic limit the contribution to the total annihilation cross-section 
times relative velocity of $\overline{\chi} \overline{\chi}$ annihilation to Higgs and 
sleptons (Fig.1) is given by: 
\bea
\langle \sigma_1|v_{\rm rel}|\rangle &=& \frac{1}{4\pi s} \frac{M_{\overline{\chi}}^2}{M_S^2}
\left( 1+ v_{\rm rel}^2/2 \right)
\left[f^2\left(1- \frac{2 M_{H}^2}{s}\right) \right. \nonumber\\
&& \left. +h_i^2\left(1-\frac{M_{\tilde{L}_{i}}^2}{s}-\frac{M_{H}^2}{s}\right)\right]\,.
\eea
\begin{figure}[htbp]
\begin{center}
\epsfig{file=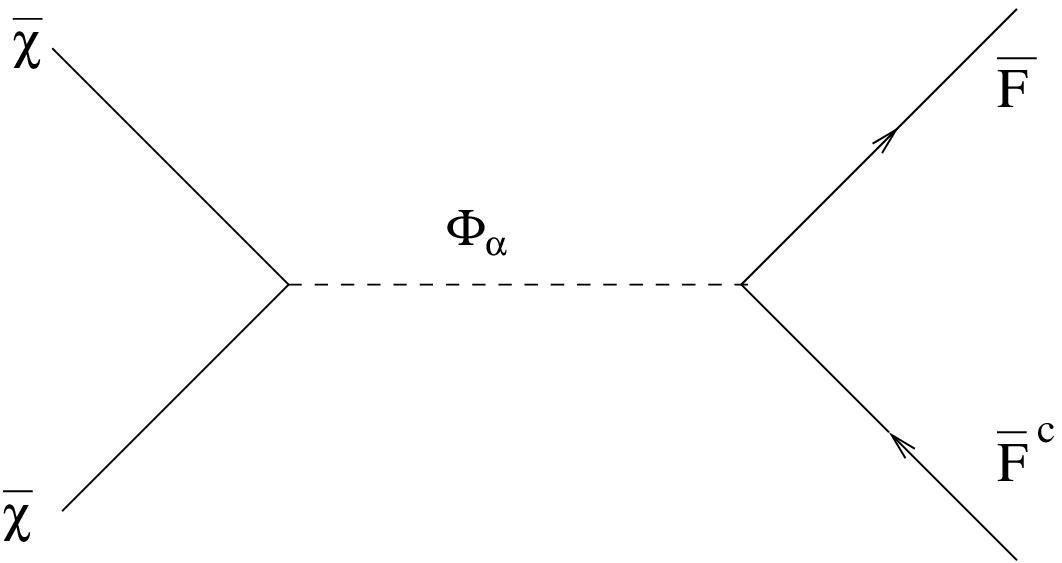, width=0.32\textwidth}
\caption{Mutual annihilation of $\overline{\chi}$ to SM fermions through Higgs and slepton. 
Here $\Phi_\alpha$=$H^{0}_u$, $H^{0}_d$, $\tilde{L}^{0}_i$, $\overline{F}$=$\overline{Q}_i$, $\overline{L}_i$, 
$\overline{H}_d$ and $\overline{F}^c$=$\overline{u}_j^c, \overline{d}_j^c, \overline{l}_j^c$.}
\label{fig2}  
\end{center}
\end{figure}
The contribution of $\overline{\chi} \overline{\chi}$ 
to SM fermions through $R$-parity conserving interactions (Fig.2) is given by
\bea
\langle \sigma_2|v_{\rm rel}|\rangle &=& \frac{1}{4 \pi s} \left( \frac{ M_{\overline{\chi}}^2}{s} 
\right) \left(1+ v_{\rm rel}^2/2 \right) 
\left[ \left(\frac{ f \langle H_d \rangle}{M_S}\right)^2 |h_{ij}^u|^2\right. \nonumber\\
&&\left. \frac{ \left(1-\frac{2 M_{\overline{u}}^{2} }{s} \right)^2}{\left( 1-\frac{M_{H}^2}{s}
\right)^2} + \left(\frac{f \langle H_u \rangle}{M_S}\right)^2 |h_{ij}^d|^2 \frac{\left(1-\frac{2 
M_{\overline{d}}^2}{s} \right)^2}{\left( 1-\frac{M_{H}^2}{s}\right)^2} \right.\nonumber\\
&& \left.+ \left(\frac{ f \langle H_u \rangle}{M_S}\right)^2 |h_{ij}^e|^2
\frac{\left(1-\frac{2 M_{\overline{L}}^2}{s}\right)^2}{\left( 1-\frac{M_{H}^2}{s}\right)^2} 
\right]\,.
\eea
The contribution of $\overline{\chi} \overline{\chi}$ 
to SM fields through $R$-parity violating interactions (Fig.2) is given by 
\bean 
\langle \sigma_3|v_{\rm rel}|\rangle &=& \frac{1}{4 \pi s} \left( \frac{ \Mchitilde^2}{s} \right) 
\left(1 + v_{\rm rel}^2/2\right)
 \nonumber\\
&& \left[  \left(\frac{h_i\langle H_u \rangle}{M_S}\right)^2 
|h_{ij}^e|^2 \frac{\left( 1- \frac{M_{\overline{L}}^2}{s} - \frac{M_{\overline{H}_{d}}^2}{s} 
\right)^2} {\left(1-\frac{\MLtilde^2}{s}\right)^2}\right.\nonumber\\
&& \left. + \left( \frac{h_i\langle H_u \rangle}{M_S}\right)^2 |\lambda_{ijk}|^2 
\frac{ \left(1-\frac{2 M_{\overline{L}}^2}{s} \right)^2}{\left( 1-\frac{\MLtilde^2}{s} 
\right)^2} \right.\nonumber\\
&& \left. + \left( \frac{h_i\langle H_u \rangle}{M_S}\right)^2 |\lambda^{'}_{ijk}|^2
\frac{ \left(1-\frac{2 M_{\overline{u}}^2}{s} \right)^2}{\left( 1-\frac{\MLtilde^2}{s} 
\right)^2}\right]
\eean
\be{n1}    ~\ee

\begin{figure}[htbp]
\begin{center}
\epsfig{file=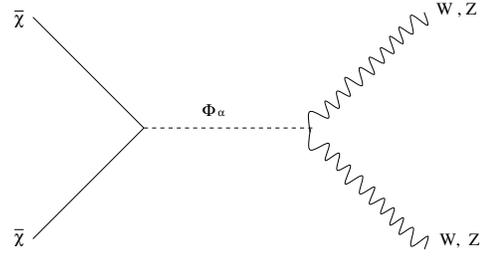, width=0.35\textwidth}
\caption{Mutual annihilation of $\overline{\chi}\overline{\chi}$ to gauge bosons through 
Higgs and sleptons. Here $\Phi_\alpha=H^{0}_u, H^{0}_d$.}
\label{fig3}
\end{center}
\end{figure}

The contribution of $\overline{\chi}\overline{\chi}$ 
to $W$-bosons (Fig.3) is given by 
\bea
\langle \sigma_4|v_{\rm rel}|\rangle &=&  \frac{1}{4 \pi} \frac{\Mchitilde^2}{s} 
\left( 1+ v_{rel}^2/2 \right) \left(2+\frac{(s-2M_W^2)^2}{4M_W^4}\right) \nonumber\\
&& \left( 1-\frac{2 M_W^2}{s}\right) \left[ \frac{ \left( \frac{f \langle H_d \rangle}{M_S} 
\right)^2 \langle H_u \rangle^2 \left( \frac{2 M_W^2}{v^2} \right)^2}{(s-M_{H}^2)^2} 
\right. \nonumber\\
&&\left. +\frac{ \left( \frac{f \langle H_u \rangle}{M_S} \right)^2 \langle H_d \rangle^2 
\left( \frac{2 M_W^2}{v^2} \right)^2}{(s-M_{H}^2)^2} \right]\, ,
\eea
where $v \equiv (\langle H_{u} \rangle^{2} + \langle H_{d} \rangle^{2})^{1/2} = 246$ GeV, while the contribution of $\overline{\chi}\overline{\chi}$ to 
$Z$-bosons (Fig.3) is given by
\bea
\langle \sigma_5|v_{\rm rel}|\rangle &=&  \frac{1}{8 \pi} \frac{\Mchitilde^2}{s}\left( 1 
+  v_{\rm rel}^2/2 \right) \left(2+\frac{(s-2M_Z^2)^2}{4M_Z^4}\right) \nonumber\\
&& \left( 1-\frac{2 M_Z^2}{s}\right) \left[ \frac{ \left( \frac{ f^2 \langle H_d \rangle}{M_S}
\right)^2 \langle H_u \rangle^2 \left( \frac{2 M_Z^2}{v^2} \right)^2}{(s-M_{H}^2)^2} 
\right. \nonumber\\
&&\left. +\frac{ \left( \frac{ f^2 \langle H_u \rangle}{M_S} \right)^2 \langle H_d \rangle^2 
\left( \frac{2 M_Z^2}{v^2} \right)^2}{(s-M_{H}^2)^2} \right]\,.
\eea
\begin{figure}[htbp]
\begin{center}
\epsfig{file=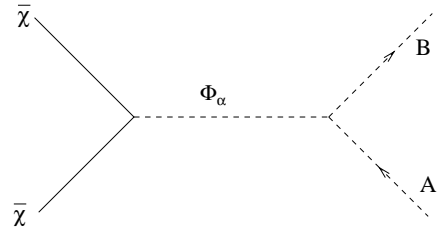, width=0.32\textwidth}
\caption{Mutual annihilation of $\overline{\chi}\overline{\chi}$ to sparticles and Higgs. 
Here $\Phi_\alpha=H^{0}_u, H^{0}_d, \tilde{L}^{0}_i$ and A, B stands for the sparticles and 
Higgs.}
\label{fig4}
\end{center}
\end{figure}
The contribution of $\overline{\chi}\overline{\chi}$ to sparticles and MSSM Higgs bosons 
(Fig.4) is given by
\bean
\langle \sigma_6|v_{\rm rel}|\rangle &=& \frac{1}{4 \pi} \frac{\Mchitilde^2}{s} \left( 1 + 
v_{\rm rel}^2/2 \right)\nonumber\\
&&\left[\frac{ \left(\frac{f \langle H_d \rangle}{M_S}\right)^2}{(s-M_{H}^2)^2}\sum_{AB}
|{\cal M}_{AB}|^2 \left(1-\frac{M_A^2}{s}-\frac{M_B^2}{s}\right) \right.\nonumber\\
&&\left. + \frac{ \left(\frac{f \langle H_u \rangle}{M_S}\right)^2}{(s-M_{H}^2)^2}
\sum_{AB}|{\cal M}_{AB}|^2 \left(1-\frac{M_A^2}{s}-\frac{M_B^2}{s}\right) \right.\nonumber\\
&&\left. +\frac{ \left(\frac{ h_i \langle H_u \rangle}{M_S}\right)^2}{(s-\MLtilde^2)^2}\sum_{AB}|
{\cal M}_{AB}|^2 \left(1-\frac{M_A^2}{s}-\frac{M_B^2}{s}\right) \right]\,,
\eean
\be{n2}    ~\ee
where ${\cal M}_{AB}$ is the mass dimension coupling at the tri-linear scalar vertex. 
\begin{figure}[htbp]
\begin{center}
\epsfig{file=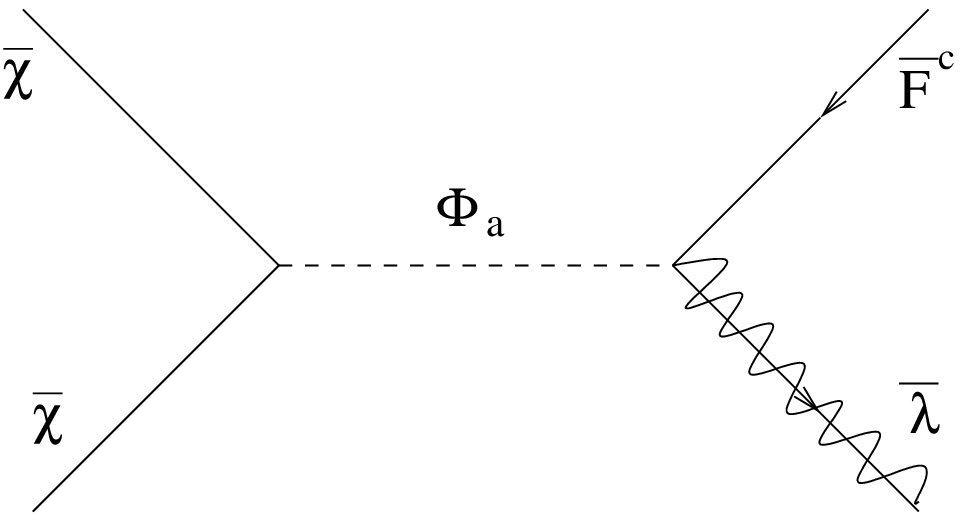, width=0.35\textwidth}
\caption{Annihilation of $\overline{\chi} \overline{\chi}$ to gauginos and fermions. 
Here $\Phi_\alpha=H^{0}_u, H^{0}_d, \tilde{L}^{0}_i$, $\overline{F}=\overline{H_u},\overline{H_d},
\overline{L}_i$ and $\overline{\lambda}=\overline{W}$, $\overline{B}$. 
\label{gaugino_gate}}
\end{center}
\end{figure}
Finally, the contribution of $\overline{\chi}\overline{\chi}$ to gaugino 
and fermion (Fig.5) is given by:
\bean
\langle \sigma_7|v_{\rm rel}|\rangle &=& \frac{1}{4 \pi s} \left( \frac{M_{\overline{\chi}}^2}{s} \right)
\left(1+ v_{\rm rel}^2/2 \right)\frac{\left(g^2+g'^2\right)}{2} 
 \times \nonumber\\
&& \left[  \left(\frac{ f \langle H_d \rangle}{M_S}\right)^2 \frac{ \left(1-\frac{M_{\overline{H}_{u}}^2}{s}-\frac{M_{\overline{\lambda}}^2}{s} \right)^2}{\left(1 - 
\frac{M_{H}^2}{s}\right)^2}  \right.\nonumber\\ 
&& \left. 
+ \left(\frac{ f\langle H_u \rangle}{M_S}\right)^2 
\frac{\left(1-\frac{M_{\overline{H}_{d}}^2}{s}-\frac{M_{\overline{\lambda}}^2}{s} \right)^2}{\left( 1-\frac{M_{H}^2}{s}
\right)^2} \right.\nonumber\\
&& \left. 
+ \left(\frac{ h_{i}\langle H_u \rangle}{M_S}\right)^2 
\frac{\left(1-\frac{M_{\overline{L}_{i}}^2}{s}-\frac{M_{\overline{\lambda}}^2}{s} \right)^2}{\left( 1-\frac{M_{\tilde{L}_{i}}^2}{s}
\right)^2}
\right]\,. 
\eean
\be{n3}        ~\ee

\section{Constraints on $R$-parity violating Interactions and implications for 
$\overline{\chi}$ annihilation}

Before estimating the relic abundance of $\overline{\chi}$ let us briefly discuss the 
constriants on $R$-parity violating interactions (\ref{rp-sup}) in the MSSM~\cite{
gautam_review}. In MSSM there are three types of tri-linear $R$-parity violating couplings: 
$\lambda_{ijk}$, $\lambda^{'}_{ijk}$ and $\lambda^{''}_{ijk}$. While $\lambda_{ijk}$ 
is antisymmetric with respect to $i$ and $j$, $\lambda^{''}_{ijk}$ is antisymmetric 
with respect to $j$ and $k$. Thus the $R$-parity violating interactions in general 
add 45 extra parameters to the MSSM. These couplings are severly constrained by the 
non-observation of certain physical phenomena. In particular, the product $\lambda^{'}
\lambda^{''}<10^{-9}$ comes from the stability of proton. Similarly, non-observation 
of $n-\bar{n}$ oscillations gives the constraint $\lambda^{''}
\leq 10^{-5}$ for $\tilde{m}=100$ GeV, where $\tilde{m}$ is the SUSY breaking mass. 
The $\lambda$ and $\lambda^{'}$ couplings induce a Majorana mass for three generations 
of neutrinos. The electron neutrino mass then gives 
the constraint $\lambda, \lambda^{'} \leq 10^{-3}$ for $\tilde{m}=100$ 
GeV. Neutrinoless double beta decay gives the constraint $\lambda^{'}\leq 10^{-4}$. Thus we see that these trilinear couplings are necessarily 
small in comparison to $R$-parity conserving couplings in the MSSM. Therefore, the annihilation 
channels of $\overline{\chi}\overline{\chi}$ through these trilinear $R$-parity violating 
couplings are necessarily small in comparison to the $R$-parity conserving couplings.

There is a bilinear term $\mu_i^{'} L_i H_u$ in the $R$-parity breaking superpotential. 
However, one can show that by making 
a $SU(4)$ rotation $\mu_i^{'} L_i H_u$ can be rotated away~\cite{hall_suzuki}, leaving 
only the bilinear term $\mu H_u H_d$ which is $R$-parity conserving. Therefore the 
presence of such a bilinear term in the $R$-parity breaking superpotential 
does not contribute to any extra annihilations of $\overline{\chi}$.  

In what follows we neglect all annihilation channels of $\overline{\chi}\overline{\chi}$ 
to MSSM fields involving $R$-parity violating couplings $\lambda,\;\lambda^{'}$ 
and $\lambda^{''}$. 
However, we note that the new $R$-parity violating couplings $h_i$ are not necessarily
small. When estimating the relic abundance of $\overline{\chi}$ we will consider only those $R$-parity 
violating channels involving the couplings $h_i$.

\section{Density of $Z_{2}$-Singlino Dark Matter}
The relic abundance of $\overline{\chi}$ can be calculated
by solving the Boltzmann equation:
\be{b1}
\frac{d\nchitilde}{dt}+3 \nchitilde H=-\langle \sigma_{\rm ann}|v_{rel}| \rangle 
\left( \nchitilde^2-{\nchitilde^{\rm eq}}^2 \right)\,,
\ee
where $\langle \sigma_{ann} |v_{rel}|\rangle $ is the thermal average of the $\overline{\chi} 
\overline{\chi}$ annihilation cross-section times relative velocity, with $\sigma_{ann}=\sum_i\sigma_i, 
i=1,7$, and $\nchitilde$ is the number density of $\overline{\chi}$.
The equilibrium density of non-relativistic $\overline{\chi}$ particles is 
\be{j2}   \nchitilde^{\rm eq} = 2 \left[ \frac{M_{\overline{\chi}}T}{2 \pi} \right]^{3/2} 
e^{-M_{\overline{\chi}}/T}    ~.\ee
With $f = \nchitilde/T^3$, \eq{b1} becomes 
\be{b2}     \frac{df}{dT} = \frac{ \langle  \sigma_{\rm ann}|v_{rel}| \rangle }{\overline{K}} \left(f^{2} - f_{eq}^{2}\right)   ~,\ee
where $f_{eq} = \nchitilde^{\rm eq}/T^3$ and $\overline{K} = \left[4 \pi^{3} g(T)/45 M_{Pl}^{2}\right]^{1/2}$. 
The density can then be calculated using the Lee-Weinberg approximation \cite{lw}.  
The freeze-out temperature, $T_{D}$, is defined by 
\be{j3}  \frac{d f_{eq}}{d T} = \frac{ \langle  \sigma_{\rm ann}|v_{rel}| \rangle }{\overline{K}} f_{eq}^{2}  ~.
\ee
To obtain the present density \eq{b2} is solved from $T_{D}$ to the present with $f_{eq} = 0$ on the right-hand side and with $f(T_{D}) = f_{eq}(T_{D})$. The freeze-out temperature can be described by 
a dimensionless parameter $z_{D} = M_{\overline{\chi}}/T_{D}$. Solving \eq{j3} gives for $z_{D}$,
\begin{equation}
z_D \equiv \frac{\Mchitilde}{T_D}  =  \ln \left[0.076\frac{1}{g_*^{1/2}}
\frac{\Mchitilde M_{\rm Pl} \langle \sigma_{\rm ann}|v_{rel}| \rangle  }{z_{D}^{1/2} \left(1 
- \frac{3}{2z_{D}}\right)} \right] \,,
\end{equation}
where 
$g_* \equiv g(T_{D})$ is the effective number
of relativsitic degrees of freedom at $T_{D}$. 
This implies that $z_D \approx 25$. Solving \eq{b2} with 
$f_{eq} = 0$ on the right-hand side and with $f(T_D) = f_{eq}(T_D)$ then gives the number density at a lower temperature,
\be{j5} n_{\overline{\chi}}(T) = \frac{g(T)}{g_{*}} \times \frac{1.67 g_{*}^{1/2}T^{3}z_{D}}{M_{\overline{\chi}} M_{Pl} 
\langle \sigma_{\rm ann}|v_{rel}| \rangle }
\frac{\left(1- \frac{3}{2 z_{D}}\right)  }{ \left(1- \frac{1}{2 z_{D}}\right) } \;\;\;;\;\; 
T \ll T_{D}         ~,\ee   
where we have included a correction for the change in the effective number of relativistic degreees 
of freedom. Therefore the present contribution of $\overline{\chi}$ to the critical density of the universe is  
\be{j6}
\omegachitilde h^2 \approx  1.1\times 10^9 {\rm GeV}^{-1} \frac{z_D}{ g_*^{1/2}M_{\rm Pl} 
\langle \sigma_{\rm ann}|v_{rel}| \rangle}
~,\ee
where $z_{D} \gg 1$ is assumed. 

In the following we consider  $\Omega_{\overline{\chi}}$ in the limits (i) $s< M_{H}^2, M_{\tilde{L}}^2$, 
and (ii) $s> M_{H}^2$, $M_{\tilde{L}}^2$, where $s\simeq 4\Mchitilde^2$ in the nonrelativistic limit.  

\subsection*{(i) Small $M_{\overline{\chi}}$: $s < M_{H}^2, M_{\tilde{L}}^2$}

\begin{figure}[htb]
\begin{center}
\epsfig{file=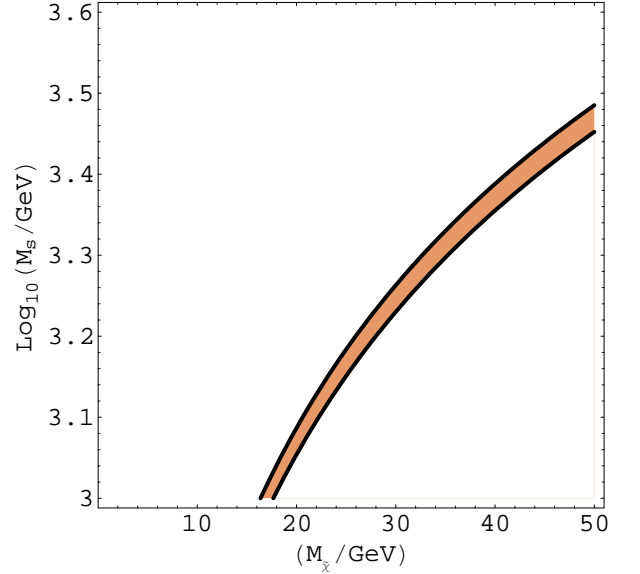, width=0.45\textwidth}
\caption{Contour of $\omegachitilde h^2=0.106\pm 0.008$ is shown in the plane 
of $M_S$ versus $\Mchitilde$. We have taken $f=h_i=1$. 
\label{fig6}} 
\end{center}
\end{figure}

  To focus on a definite example we set $M_{H} = M_{\tilde{L}} = 150$ GeV and $\tan \beta  \equiv \langle H_u \rangle / \langle H_d \rangle = 1$  in the cross-sections. We assume that the mass of the other sparticles is 100 GeV. 
Since we assume that $s< M_{H}^2, M_{\tilde{L}}^2$, in this case only $\sigma_2$, $\sigma_{3}$ and $\sigma_{7}$ will  contribute to the relic abundance 
of $\overline{\chi}$. 
We put $f=h_i=1$; the results for smaller values can be obtained by rescaling $M_{S}$.
 In this case 
the allowed region in the plane of $M_S$ versus $\Mchitilde$ for $\omegachitilde h^2=0.106\pm 0.008$ 
is shown in Fig. (\ref{fig6}).
It can be seen that for $15 \GeV \lsim \Mchitilde \lsim  50$ GeV, $M_S$ is in the range 1-3 TeV. 
The behaviour can be understood as follows. 
In the limit $s < M_{H}^{2},\MLtilde^2 $, the annihilation cross-section 
$\sigma_{\rm ann}=\sigma_2+\sigma_3+\sigma_7$ times relative velocity is of the form: 
\begin{equation}
\langle \sigma_{\rm ann} |v_{rel}| \rangle   \propto C \frac{M_{\overline{\chi}}^2}{M_S^2}  \, ,
\end{equation}
where $C$ is a dimensionful constant involving the VEV of $H_u$ and $H_d$. Therefore, smaller values 
of $M_{\overline{\chi}}$ require small values of $M_S$ in order to keep $\omegachitilde h^2$ 
constant. 

\subsection*{(ii) Large $M_{\overline{\chi}}$: $s> M_{H}^2, M_{\tilde{L}}^{2}$}
\begin{figure}[htbp]
\begin{center}
\epsfig{file=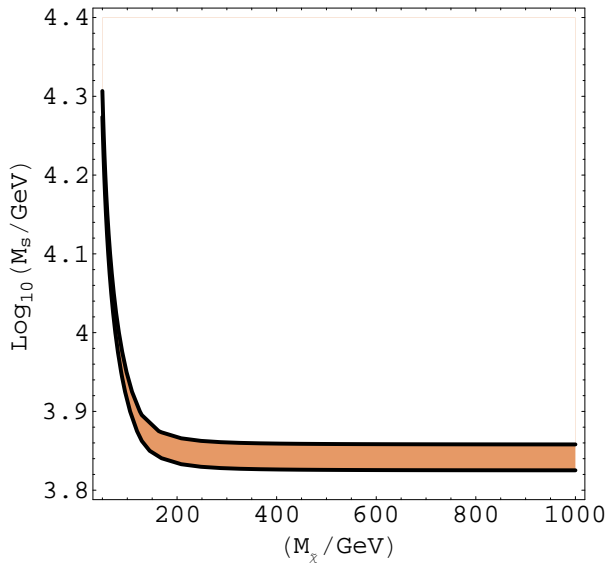, width=0.45\textwidth}
\caption{Allowed region of $\omegachitilde h^2=0.106\pm 0.008$ is shown in the plane
of $M_S$ versus $\Mchitilde$. We have taken $f=h_i=1$. 
\label{fig7}}
\end{center}
\end{figure}

We next consider $s>M_{H}^2, M_{\tilde{L}}^2$.
We show the allowed region in the plane of $M_S$ versus $\Mchitilde$, corresponding to $\omegachitilde 
h^2=0.106\pm 0.008$, in Fig. (\ref{fig7}). From  Fig. (\ref{fig7}) it can be seen that for $\Mchitilde 
\gsim 200$ GeV, $M_S$ is almost constant at around $10^{3.84} \GeV \equiv 6.9 \TeV$. 
This can be understood as follows. In the 
limit $s> M_{H}^2, \MLtilde^2$, the annihilation cross-section $\sigma_{\rm ann}=
\sum_i \sigma_i (i=1-7$) times relative velocity is of the form: 
\begin{equation}
\langle \sigma_{\rm ann}|v_{rel}| \rangle \propto \frac{1}{M_S^2}+ C \left( \frac{1}{M_{\overline{\chi}}^2 M_S^2} \right) \,,
\end{equation}
where $C$ is a dimensionful constant. 
For $\Mchitilde \gsim 200$ GeV, the effecive 
annihilation cross-section is dominated by the first term. As a result we get a constant value $M_S \approx 6.9$ TeV.
For $M_{\overline{\chi}} \lsim 200$ GeV, the second term 
in the above equation dominates. In this regime, larger $M_S$ is required to keep $\omegachitilde h^2$ constant as $M_{\overline{\chi}}$ decreases, with a Higgs pole at $M_{\overline{\chi}} = 75$ GeV  allowing much larger values of $M_{S}$ over a small range of $M_{\overline{\chi}}$.

In general, smaller values 
of $M_{S}$ are possible by reducing $f$ and $h_{i}$, so the values shown in the figures should be considered as 
upper bounds on $M_{S}$, corresponding to large $f$ and $h_{i}$. 

\section{Conclusions and Outlook}
We have discussed the possibility of $Z_2$-singlino dark matter 
in extensions of the MSSM. The dark matter communicates with 
visible matter through a heavy messenger field, $S$. As a result the interaction is suppressed by the mass scale $M_S$. 
For $M_S\lsim 10^4$ GeV the $Z_{2}$-singlino can be cold dark matter for a wide range 
of mass, $15 {\rm GeV} \lsim \Mchitilde \lsim 1 {\rm TeV}$. (Larger values of $M_{S}$ are possible near a Higgs pole.) 
The possibility of dark matter in this case does not rely on the conservation of $R$-parity. Thus the model is particularly important for the MSSM and its extensions, such as the  
NMSSM and nMSSM, when $R$-parity is violated. Non-conservation of $R$-parity is often considered to give small neutrino masses, as required by the oscillation 
data, and for leptogenesis, a robust mechanism for the matter anti-matter asymmetry of the Universe. 

In the case of non-SUSY gauge singlet scalars interacting via the Higgs portal, direct and 
indirect detection rates are comparable with conventional weakly interacting dark matter 
candidates \cite{singlets,singlet2}. In the $Z_{2}$-singlino case the coupling to the Higgs has 
an additional suppression factor $\approx v/M_{S}$, where $v$ is a Higgs expectation value. Therefore 
we would expect significant detection rates for $M_{S} \; ^{<}_{\sim} \;  1$ TeV. In this case the 
effective theory based on integrating out the $S$ fields may not be appropriate. We will return to the 
question of $Z_{2}$-singlino detection in future work.

      The $Z_2$ symmetry responsible for dark matter in this model can be a surviving symmetry (a 
discrete gauge symmetry) of a gauged $U(1)'$ extension of MSSM. Such models are natural 
in top-down scenarios when $E(6)$ grand unified theory is broken down to the MSSM. A gauge origin of the 
$Z_{2}$ is favoured by arguments which suggest that global symmetries, both continuous and discrete, 
are broken by non-perturbative gravitational effects \cite{coleman}. In this case $R$-parity may be broken 
while a $Z_{2}$ discrete gauge symmetry may account for SUSY dark matter.

    We have focused on the case of $Z_{2}$-singlino dark matter produced by conventional freeze-out from 
thermal equilibrium. There is, however, another possibility. In the case of non-SUSY gauge singlet scalar dark 
matter, when the mass of the scalar is entirely generated by the Higgs expectation value, the correct relic 
density is produced via decay of thermal background Higgs bosons when the mass of singlet scalars is in the range 1-10 
MeV \cite{lights}. This is the ideal range \cite{int1,int2} for very long-lived dark matter particles to account 
for the 511 keV line obeserved by INTEGRAL \cite{integral}. In the $Z_{2}$-singlino model, the singlino mass will 
be entirely generated by the Higgs expectation value in the limit $M_{\chi} \rightarrow 0$. We will consider the 
light $Z_{2}$-singlino in a forthcoming paper \cite{sahu_pre}.    
 
Although we have considered dark matter particles which are Standard Model singlets, the model can 
easily be generalised, for example, to a SUSY version of the inert doublet dark matter model \cite{inert}. In 
addition, the messenger mass in the model can be greater than $10^{4}$ GeV, in particular for the case where the $Z_{2}$-singlino mass is close to a Higgs pole. This may allow the messengers to be associated with the messenger fields of a gauge mediated SUSY 
breaking model. 

              The model we have presented here may be regarded as a SUSY generalisation of the Higgs portal 
concept. As such, we can expect the model to arise in the low energy effective theory of a wide range of SUSY 
particle physics models.  

\noindent {\bf Comment:} While this paper was in preparation a similar model was presented in \cite{mretal}.  

\section*{Acknowledgement}
NS would like to thank Ki Young Choi and Kazunori Kohri for useful discussions. JM and NS were supported 
by the European Union through the Marie Curie Research and Training Network "UniverseNet" (MRTN-CT-2006-035863) 
and by STFC (PPARC) Grant PP/D000394/1.

\end{document}